\documentclass[11pt]{article}
\usepackage{amsfonts}
\usepackage[nosort]{cite}
\csname @addtoreset\endcsname{equation}{section}
\newcommand{\fft}[2]{{\frac{#1}{#2}}}
\newcommand{\ft}[2]{{\textstyle{\frac{#1}{#2}}}}
\newcommand{\sqr}[2]{{\vcenter{\vbox{\hrule height.#2pt
        \hbox{\vrule width.#2pt height#1pt \kern#1pt
        \vrule width.#2pt}\hrule height.#2pt}}}}

\newcommand{\be}{\begin{equation}}
\newcommand{\ee}{\end{equation}}
\newcommand{\bea}{\begin{eqnarray}}
\newcommand{\eea}{\end{eqnarray}}
\begin{document}
%%%%%%%%%%%%%%%%%%%%%%%%%%%%%%%%%%%%%%%%%%%%%%%%%%%%%%%%%%%%%%%%%%%%%%%%%%%%%%
\begin{titlepage}

\hbox to \hsize{{\tt hep-th/0412242}\hss{\tt MCTP-04-72}}

\vskip 1.7 cm

\begin{center}
{\bf \Large 
Bubbling 1/2 BPS solutions of minimal six-dimensional supergravity}

\vskip 1cm

{\bf \large James T. Liu and Diana Vaman}  

\vskip 1cm
{\it Michigan Center for Theoretical Physics,
Randall Laboratory of Physics,\\
The University of Michigan, Ann Arbor, MI 48109--1120}
\vskip .5 cm
{E-mail: \{{\tt jimliu,dvaman}\}\tt@umich.edu}  

\end{center}

\vspace{1cm}

\begin{abstract}

We continue our previous analysis (hep-th/0412043) of 1/2~BPS solutions
to minimal 6d supergravity of bubbling form. We show that, by turning on an
axion field in the $T^2$ torus reduction, the constraint 
$F\wedge F$, present in the case of an $S^1\times S^1$ reduction, 
is relaxed. We prove that the four-dimensional reduction to a bosonic
field theory,
whose content is the metric, a gauge field, two scalars and a 
pseudo-scalar (the axion), is consistent. Moreover, these reductions
when lifted to the six-dimensional minimal supergravity represent the 
sought-after family of 1/2~BPS bubbling solutions.

\end{abstract}

\end{titlepage}

%%%%%%%%%%%%%%%%%%%%%%%%%%%%%%%%%%%%%%%%%%%%%%%%%%%%%%%%%%%%%%%%%%%%%%%%%%%%%%
\section{Introduction}

In this note we complete the search for bubbling 1/2~BPS solutions of
minimal six dimensional supergravity initiated in \cite{Liu:2004ru}.  These
configurations were in turn motivated by \cite{Lin:2004nb}, where
1/2~BPS supergravity solutions corresponding to bubbling-type deformations
of AdS$_5 \times S^5$ geometry were shown to admit a dual description in
terms of free fermions \cite{Berenstein:2004kk,Lin:2004nb}. 

In \cite{Liu:2004ru}, inspired by \cite{Lin:2004nb}, we considered solutions
of minimal six dimensional supergravity, which had an $S^1\times S^1$ isometry
$ds^2=g_{\mu\nu} dx^\mu dx^\nu+e^{H(x)+G(x)} d\phi_1^2 + e^{H(x)-G(x)}
d\phi_2^2$, and $-2 H_{(3)}=F_{(2)}d\phi_1+\widetilde F_{(2)} d\phi_2$. 
We found that this metric ansatz did not allow the existence 
a family of 1/2~BPS solutions, because of an additional constraint 
$F_{(2)}\wedge F_{(2)}=0$. We showed that this constraint arises for all
metric reductions of the type $S^n\times S^n$, with the exception of the case
$n=3$, which was the case in \cite{Lin:2004nb}. From a bosonic reduction
perspective, the $S^1\times S^1$ reduction was inconsistent; it had 
to be supplemented by hand by the above mentioned constraint. Otherwise said,
a particular six dimensional equation of motion, the Einstein equation with
components $\phi_1\phi_2$ could not be recovered from the Lagrangian 
of the effective four dimensional bosonic theory, and it corresponded precisely
to the constraint $F_{(2)}\wedge F_{(2)}=0$. Therefore, we conjectured that
by turning on an axion field, this constraint could in principle be 
eliminated. Interestingly enough, the AdS$_3\times S^3$, the maximally 
symmetric plane wave, and the multi-center string arising from a D1-D5
configuration were shown to satisfy the constraint.

Here we demonstrate that, indeed, a generic $T^2$ torus reduction ansatz 
\bea
ds^2&=&g_{\mu\nu}dx^\mu dx^\nu + e^{H(x)+G(x)} d\phi_1^2 + e^{H(x)-G(x)}
\left(d\phi_2 + \chi(x) d\phi_1\right)^2\nonumber\\
-2H_{(3)}&=&
F_{(2)}\wedge d\phi_1+\widetilde F_{(2)}\wedge (d\phi_2+\chi d\phi_1)
\label{ansatz}
\eea
not only
eliminates the need for the constraint, but at the same time leads to
the construction of the family of 1/2~BPS solutions which correspond
to bubbling-type deformations of the AdS$_3\times S^3$ geometry.   
In (\ref{ansatz}) we denoted by $\chi(x)$ the additional field to be
kept in the Kaluza-Klein truncation, which retains in the 
four dimensional bosonic effective field theory besides the metric, a 
gauge field ($F_{(2)}$ and $\widetilde F_{(2)}$ are related by the 
self-duality of the 3-form field strength $H_{(3)}$), two scalars
$H, G$, and a pseudo-scalar, the axion $\chi$. In contrast to the
conclusion drawn in \cite{Liu:2004ru}, where the rectangular torus reduction
was inconsistent, we prove that by retaining the axion, we have achieved
a consistent {\it bosonic} truncation. 
In order to be able to lift the bosonic four dimensional
solutions to a family of supersymmetric solutions which includes AdS$_3\times S^3$, the 
Killing spinors must be charged under the two $U(1)$ symmetries.

It is worth noting that, while the bubbling AdS$_5$ solutions of type IIB
supergravity were constructed in terms of a harmonic function which
had to obey certain boundary conditions in order for the solution
to be non-singular, here we find that the bubbling AdS$_3$ solution is
written in terms of two independent functions, each obeying second order
differential equations.

The paper is structured following a similar pattern to \cite{Liu:2004ru}:
in Section 2 we show that the $T^2$ reduction ansatz yields a consistent
{\it bosonic} reduction of minimal six dimensional supergravity and
furthermore perform a reduction of the gravitino supersymmetry variation. 
In Section 3 we construct the Killing spinor associated with this $T^2$
reduction, and obtain the sought-after family of 1/2~BPS solutions
corresponding to a bubbling AdS$_3$.  We conclude with a discussion
section.  Finally, the appendices contain the analysis
of the integrability of the supersymmetry variations and the full set of 
differential and algebraic identities obeyed by the spinor bilinears.

%%%%%%%%%%%%%%%%%%%%%%%%%%%%%%%%%%%%%%%%%%%%%%%%%%%%%%%%%%%%%%%%%%%%%%%%%%%%%%
\section{Bosonic reduction of minimal $D=6$ supergravity on $T^2$}
\label{sec:s1s1}

As in \cite{Liu:2004ru}, we are concerned with the reduction of $D=6$,
${\cal N}=(1,0)$ supergravity to yield an effective theory in four
dimensions.  Unlike \cite{Liu:2004ru}, however, which focused on the
$S^1\times S^1$ reduction, we now consider the full $T^2$ reduction,
allowing in particular a mixing between the two $\mathrm U(1)$ isometries,
related to the tilting of the torus.

Although it is straightforward to couple to a tensor multiplet
(which would be necessary for more general D1-D5 systems), here we consider
only the minimal $\mathcal N=(1,0)$ supergravity, consisting of the
gravity multiplet ($g_{MN}$, $\psi_M$, $B_{MN}^+$), where
$B_{MN}^+$ denotes a two-form potential with self-dual field strength,
$H_{(3)}=dB_{(2)}^+$ and $\psi_M$ is a left-handed gravitino satisfying
the projection $\Gamma^7\psi_M=-\psi_M$.  Here we are following the
notation introduced in \cite{Liu:2004ru}.

The bosonic equations of motion for the supergravity multiplet are simply
\begin{equation}\label{eq:eom}
R_{MN}=\ft14H_{MPQ}H_N{}^{PQ},\qquad H_{(3)}=*H_{(3)},\qquad dH_{(3)}=0.
\end{equation}
Although this theory does not admit a covariant Lagrangian formulation,
we may formally take
\begin{equation}
e^{-1}{\cal L}=R-\fft1{2\cdot3!}H_{(3)}^2,
\end{equation}
with the addition that the self-duality condition on $H_{(3)}$ must be
imposed by hand after obtaining the equations of motion.

Following \cite{Lin:2004nb,Liu:2004ru}, we proceed with a nearly
standard Kaluza-Klein reduction on $T^2$, given by
\begin{eqnarray}
&&ds^2=g_{\mu\nu}(x)dx^\mu dx^\nu+e^{H(x)}\left( e^{G(x)}d\phi_1^2
+e^{-G(x)}(d\phi_2+\chi(x)d\phi_1)^2\right),\nonumber\\
&&-2H_{(3)}=F_{(2)}e^{-\fft12(H+G)}\wedge e^{\underline 4}
+\widetilde F_{(2)}e^{-\fft12(H-G)}
\wedge e^{\underline 5}.
\label{eq:ans}
\end{eqnarray}
In writing $H_{(3)}$, and in the subsequent expressions, we use the natural
vielbein basis
\begin{equation}
e^{\underline 4}= e^{\fft12(H+G)}d\phi_1,
\qquad e^{\underline 5}=e^{\fft12(H-G)}(d\phi_2 + \chi d\phi_1).
\end{equation}
Although we have written the $H_{(3)}$ ansatz in (\ref{eq:ans}) in terms
of four-dimensional gauge fields $F_{(2)}$ and $\widetilde F_{(2)}$, these
fields are not independent, but are related by the condition that $H_{(3)}$
is self-dual.  Computing
\begin{equation}
-2*H_{(3)}=*_4 \widetilde F_{(2)}e^{-\fft12(H+G)}\wedge e^{\underline 4}
-*_4 F_{(2)}e^{-\fft12(H-G)} \wedge e^{\underline 5},
\end{equation}
we see that the self-duality condition $H_{(3)}=*H_{(3)}$ implies
\be
\widetilde F_{(2)}=-e^{-G} *_4 F_{(2)},\qquad F_{(2)}=e^{G} *_4
\widetilde F_{(2)}.
\label{eq:h3sd}
\ee
Therefore, the effective bosonic reduction will result in a four-dimensional
system consisting of the metric $g_{\mu\nu}$, a gauge field $F_{(2)}$, the
two scalars $H$, $G$, and a pseudo-scalar `axion' $\chi$.

At this stage, it is worth commenting on the structure of the reduction
ansatz.  Recall that a standard $T^2$ reduction of the minimal $\mathcal N
=(1,0)$ theory would result in $\mathcal N=2$ supergravity coupled to
two vector multiplets in four dimensions.  The two vector multiplets
contain two scalars and two pseudoscalars, together parameterizing two
$\mathrm{SL}(2,\mathbb R)/\mathrm U(1)$ cosets, one related to the
complex structure of $T^2$ and the other to its K\"ahler modulus.  In
contrast, here we set both metric gauge fields as well as the axionic
scalar from the K\"ahler modulus to zero.  Thus only the complex structure
$\mathrm{SL}(2,\mathbb R)$ survives, given by the complex parameter
$\tau=\chi+ie^G$.  The remaining scalar $e^H$ parameterizes the volume
of $T^2$, but is otherwise missing its axionic partner $\widetilde\chi$
ordinarily arising from an addition to the $H_{(3)}$ reduction ansatz in
(\ref{eq:ans}) of the form $(1+*)d\widetilde\chi\wedge e^{\underline4}
\wedge e^{\underline5}$.  Nevertheless, although this reduction is
incomplete from a supersymmetric point of view (as it results in a
non-supersymmetric field content), we will see below that it is a
consistent reduction of the bosonic sector.  The addition of the complex
structure axion $\chi$ is crucial for consistency.

Proceeding with the bosonic reduction, we note that, in addition to
the self-duality condition on $H_{(3)}$, the equation of motion
$dH_{(3)}=0$ results in the form-field equations
\begin{eqnarray}
d\widetilde F_{(2)}=0,\qquad dF_{(2)}+\widetilde F_{(2)}\wedge d\chi=0.
\label{eq:afeom}
\end{eqnarray}
It is this result here that indicates that $\widetilde F_{(2)}=d\widetilde
A_{(1)}$ has a natural representation in terms of a potential, while
$F_{(2)}$ has a more complicated representation.  The form-field provides
a source to Einstein's equations.  We compute
\begin{eqnarray}
(H_{(3)}^2)_{\mu\nu}&=&\ft 12e^{-(H+G)}(F^2)_{\mu\nu}+\ft 12
e^{-(H-G)}(\widetilde F^2)_{\mu\nu},\nonumber\\
(H_{(3)}^2)_{\underline 4 \underline 4}&=&\ft 14 e^{-(H+G)}F^2,\qquad
(H_{(3)}^2)_{\underline 5\underline 5}=\ft 14e^{-(H-G)}\widetilde F^2,\nonumber\\
(H_{(3)}^2)_{\underline 4\underline 5}&=&\ft 14 e^{-H}
F_{\mu\nu}\widetilde F^{\mu\nu}.
\label{eq:h3sour}
\end{eqnarray}
Note that
\begin{equation}
H_{(3)}^2=\ft34e^{-(H+G)}F^2+\ft34e^{-(H-G)}\widetilde F^2.
\end{equation}
However, by using the self-duality condition (\ref{eq:h3sd}), we see
simply that $H_{(3)}^2=0$, which is a kinematical constraint from
self-duality.

Turning to the Einstein equations, we first compute the spin connections
\bea
&&\omega^{\underline{4}\underline{5}}=\ft 12 e^{-G} d\chi,\nonumber\\
&&\omega^{\underline{4}\underline{m}}e_{\underline{m}\mu}=
-\ft 12 \partial_\mu (H+G) e^{\underline 4}
-\ft 12 e^{-G}\partial_\mu\chi e^{\underline 5},\nonumber\\ 
&&\omega^{\underline{5}\underline{m}}e_{\underline{m}\mu}=-\ft12 
\partial_\mu (H-G)e^{\underline 5}-\ft 12 e^{-G}\partial_\mu\chi
e^{\underline 4},
\eea
as they also prove useful in the reducing the supersymmetry variations,
below.  It is then a straightforward exercise to compute the Riemann
tensor through $R=d\omega+\omega\wedge\omega$, and then the Ricci
tensor.  In frame components, we obtain
\begin{eqnarray}
R_{\mu\nu}&=&\hat R_{\mu\nu}-\ft{1}2(\partial_\mu H\partial_\nu H
+\partial_\mu G\partial_\nu G+\partial_\mu\chi\partial_\nu\chi e^{-2G})
-\nabla_\mu\nabla_\nu H,\nonumber\\
R_{\underline{4}\underline{4}}&=&-\ft{1}2\partial^\mu H\partial_\mu(H+G)
-\ft12 \square(H+G)-\ft 12\partial_\mu\chi\partial^\mu\chi e^{-2G},\nonumber\\
R_{\underline{4}\underline{5}}&=&-\ft12e^{-G}\left(\square\chi
+\partial_\mu(H-2G)\partial^\mu \chi\right),\nonumber\\
R_{\underline{5}\underline{5}}&=&-\ft{1}2
\partial^\mu H\partial_\mu(H-G)-\ft12
\square(H-G)+\ft 12\partial_\mu\chi\partial^\mu\chi e^{-2G}.
\label{eq:a4eins}
\end{eqnarray}
Combining these expressions with the source (\ref{eq:h3sour}), we obtain
the four-dimensional equations of motion
\begin{eqnarray}
\kern-16pt&&R_{\mu\nu}=\ft{1}2(\partial_\mu H\partial_\nu H+\partial_\mu G\partial_\nu G +e^{-2G}\partial_\mu\chi\partial_\nu\chi)
+\nabla_\mu\nabla_\nu H+\ft 14e^{-(H-G)}\Bigl((\widetilde F^2)_{\mu\nu}
-\ft 14 g_{\mu\nu} \widetilde F^2\Bigr),\nonumber\\
\kern-16pt&&\nabla^\mu\Big(e^H\nabla_\mu H\Big)=0,\qquad
\nabla^\mu\Big(e^H\nabla_\mu G\Big)=
-e^{H-2G}\partial_\mu\chi\partial^\mu\chi + \ft 18 e^G \widetilde F^2,
\nonumber\\
\kern-16pt&&\nabla^\mu\Big(e^{H-2G}\nabla_\mu\chi\Big)=-\ft 1{16}\epsilon_{\mu\nu\rho\sigma}
\widetilde F^{\mu\nu}\widetilde F^{\rho\sigma}.
\label{eq:aeeom}
\end{eqnarray}
The scalar equations were separated by taking appropriate linear combinations
of the $R_{\underline4\underline4}$ and $R_{\underline5\underline5}$ equations.  

We now see that the equations of motion, (\ref{eq:afeom}) and
(\ref{eq:aeeom}), may be
derived from an effective four-dimensional Lagrangian
\begin{eqnarray}
\kern-8pt
e^{-1}{\cal L}&=&e^{H}\Bigl[R+\ft12\partial H^2-\ft12\partial G^2-\ft 12
\partial\chi^2 e^{-2G}-\ft 18 e^{-(H-G)}\widetilde F^2+\ft 1{16} \chi\epsilon_{\mu\nu\rho\sigma}\widetilde F^{\mu\nu}
\widetilde F^{\rho\sigma}\Bigr].\qquad
\label{eq:aslag}
\end{eqnarray}
The inclusion of the axion extends the analysis of \cite{Liu:2004ru},
and removes the $F_{(2)}\wedge F_{(2)}=0$ constraint.  It is of
course precisely $F_{(2)}\wedge F_{(2)}$ that sources the axion, and
this is the origin of the inconsistency if the axion were to be truncated
by hand.

%%%%%%%%%%%%%%%%%%%%%%%%%%%%%%%%%%%%%%%%%%%%%%%%%%%%%%%%%%%%%%%%%%
\subsection{Supersymmetry variations}

Having completed the reduction of the bosonic sector with the axion,
we now proceed to reduce the gravitino variation
\begin{equation}
\delta\psi_M=[\nabla_M+\ft1{48}H_{NPQ}\Gamma^{NPQ}\Gamma_M]\varepsilon.
\label{eq:6gto}
\end{equation}
Much of the analysis of the fermionic sector parallels that of
\cite{Liu:2004ru}.  However, some care must be taken when working with
an off-diagonal metric on $T^2$.  Following an identical Dirac decomposition
\begin{equation}
\Gamma_\mu=\gamma_\mu\times\sigma_1,\qquad
\Gamma_{\underline 4}=1\times\sigma_2,\qquad
\Gamma_{\underline 5}=\gamma_5\times\sigma_1,
\end{equation}
as well as the projection conditions $\Gamma^7\varepsilon=
-\varepsilon$ and $\Gamma^7\psi_M=-\psi_M$ on Weyl spinors, the
six-dimensional gravitino variation becomes
\begin{eqnarray}
\delta\psi_{\mu}&=&[\nabla_{\mu}-\ft i4e^{-G}\partial_\mu\chi \gamma_5 
+\ft{i}{16}e^{-\ft{1}{2}(H+G)}F_{\nu\lambda}\gamma^{\nu\lambda}\gamma_{\mu}]\epsilon,
\nonumber\\
\delta\lambda_H&=&[\gamma^{\mu}\partial_{\mu}H+2i e^{-\fft12H}
(e^{-\fft12G}(\partial_{\phi_1}-\chi\partial_{\phi_2})-i\gamma_5 e^{\fft12G}\partial_{\phi_2})]\epsilon,\nonumber\\
\delta\lambda_G&=&[\gamma^{\mu}\partial_{\mu}G
+ie^{-G}\partial_\mu\chi\gamma_5\gamma^\mu
-\ft{i}4e^{-\fft12(H+G)}F_{\mu\nu}\gamma^{\mu\nu}\nonumber\\
&&\qquad\qquad+2ie^{-\fft12H}
(e^{-\fft12G}(\partial_{\phi_1}-\chi\partial_{\phi_2}) +i\gamma_5 e^{\fft12G}\partial_{\phi_2})]\epsilon,
\label{eq:6to4susy}
\end{eqnarray}
where we have defined the linear combinations
\begin{equation}
\lambda_H=2(i\psi_{\underline 4}+\gamma_5\psi_{\underline 5}),\qquad
\lambda_G=2(i\psi_{\underline4}-\gamma_5\psi_{\underline 5}).
\end{equation}
(Note that these spinors were defined as $\chi_H$ and $\chi_G$ in
\cite{Liu:2004ru}; here we use $\lambda_H$ and $\lambda_G$ to avoid
confusion with the axion.)  The four-dimensional Dirac spinor $\epsilon$
was related to the left-handed six-dimensional spinor by
$\varepsilon=\epsilon\times\left[1\atop0\right]$.

As highlighted in \cite{Liu:2004ru}, to obtain a bubbling ansatz, we
must allow for $\mathrm U(1)\times\mathrm U(1)$ charged Killing spinors.
Thus we write
\begin{equation}
\epsilon(x,\phi_1,\phi_2)=e^{-\fft{i}2(\eta\phi_1+\tilde\eta\phi_2)}
\epsilon(x),
\end{equation}
where the Kaluza-Klein momenta (or chargers) $\eta$ and $\tilde\eta$ are 
quantized in integer units.  This quantization is enforced by the
periodicity of the two-torus, even in the tilted case.  In the
$\mathrm{SL}(2,\mathbb Z)$ point of view, the spinor charges
$(\eta,\tilde\eta)$ transform as a doublet.  The result of using a
charged spinor is that we may make a simple replacement
\begin{equation}
\partial_\phi\to -i\ft{\eta}{2},\qquad\partial_{\tilde\phi}\to
-i\ft{\tilde\eta}{2},
\end{equation}
in the supersymmetry variations (\ref{eq:6to4susy}).

We thus see that, compared to the $S^1\times S^1$ case, the effect of
working with $T^2$ is to introduce an axion $\chi$ (corresponding to
an off-diagonal metric component $g_{\phi_1\phi_2}$) in both
(\ref{eq:aslag}) and (\ref{eq:6to4susy}).  Furthermore, the originally
independent $\mathrm U(1)$ charges $\eta$ and $\tilde\eta$ now combine
into an $\mathrm{SL}(2,\mathbb Z)$ doublet.

%%%%%%%%%%%%%%%%%%%%%%%%%%%%%%%%%%%%%%%%%%%%%%%%%%%%%%%%%%%%%%%%%%%%%%%%%%%%%%
\section{The bubbling AdS$_3$ solution}

We now complete the supersymmetry analysis in the presence of the axion
$\chi$.  Following \cite{Lin:2004nb,Liu:2004ru}, we introduce the spinor
bilinears
\begin{eqnarray}
&&f_1=\bar\epsilon\gamma^5\epsilon,\qquad f_2=i\bar\epsilon\epsilon,
\nonumber\\
&&K^\mu=\bar\epsilon\gamma^\mu\epsilon,\qquad L^\mu=\bar\epsilon\gamma^\mu
\gamma^5\epsilon,\nonumber\\
&&Y_{\mu\nu}=i\bar\epsilon\gamma_{\mu\nu}\gamma^5\epsilon,
\label{eq:bilinear}
\end{eqnarray}
where the factors of $i$ are chosen to make these tensor quantities real.
Using the methods of
\cite{Gauntlett:2002sc,Gauntlett:2002nw,GMR,Gauntlett:2004zh,Lin:2004nb,Liu:2004ru},
we proceed to examine the algebraic and differential identities satisfied
by the above tensors.  The useful algebraic identities are straightforward:
\begin{equation}
L^2=-K^2=f_1^2+f_2^2,\qquad K\cdot L=0.
\label{eq:h2}
\end{equation}
In addition, the complete set of differential identities are provided in
Appendix~\ref{sec:apc}.

We first fix the form of the scalar
quantities $f_1$ and $f_2$.  Combining the differential identities
for $\nabla_\mu f_1$ and $\nabla_\mu f_2$ in (\ref{eq:difdi}) with
the $L_\mu$ identities in (\ref{eq:difhi}) and (\ref{eq:difgi}), we
obtain
\begin{eqnarray}
\partial_\mu f_1&=&\ft14e^{-\fft{1}2(H+G)}*F_{\mu\nu}K^\nu
+\ft 12 f_2e^{-G}\partial_\mu\chi=\ft12f_1\partial_\mu(H-G)+f_2 e^{-G}\partial_\mu\chi
,\nonumber\\
\partial_\mu f_2&=&-\ft14e^{-\fft{1}2(H+G)}F_{\mu\nu}K^\nu-\ft 12 f_1 e^{-G}
\partial_\mu\chi=\ft12f_2\partial_\mu(H+G).
\label{eq:fkeqn}
\end{eqnarray}
This gives two equations for $f_1$ and $f_2$
\begin{equation}
\partial_\mu[e^{-\fft12(H-G)}f_1]=[e^{-\fft12(H+G)}f_2]\partial_\mu\chi,\qquad
\partial_\mu[e^{-\fft12(H+G)}f_2]=0,
\end{equation}
which may be integrated to obtain
\begin{equation}
f_1=(b+a\chi)e^{\fft12(H-G)},\qquad f_2=ae^{\fft12(H+G)}.
\label{eq:f1f2eqn}
\end{equation}
In addition, the constants $a$ and $b$ are related through the identity 
$(\eta-\chi\widetilde\eta) f_2=-\widetilde\eta e^Gf_1$ of (\ref{eq:difhi}).  
In particular
\begin{equation}
a\eta+b\widetilde\eta=0.
\label{eq:abeta}
\end{equation}
Comparing with the $S^1\times S^1$ compactification \cite{Liu:2004ru},
we see that at this point the only effect of the axion is to shift $f_1$
in (\ref{eq:f1f2eqn}).

Given $f_1$ and $f_2$, we may now fix the normalization of the vectors
$K_\mu$ and $L_\mu$.  Using (\ref{eq:h2}), we obtain
\begin{equation}
L^2=-K^2=f_1^2+f_2^2=e^H(a^2e^G+(b+a\chi)^2e^{-G}).
\label{eq:knorm}
\end{equation}
Furthermore, the $L_\mu$ equations of (\ref{eq:difhi}) provide the constraints
\begin{equation}
\eta L_\mu=b\partial_\mu e^H,\qquad\widetilde\eta L_\mu=-a\partial_\mu e^H,
\label{eq:diffh}
\end{equation}
which are axion independent.

Following \cite{Lin:2004nb}, we now observe from (\ref{eq:difdi}) that
both $K_{(\mu;\nu)}=0$ so that $K^\mu$ is a Killing vector and $dL=0$.
We thus choose a preferred coordinate basis
so that the Killing vector $K^\mu\partial_\mu$ corresponds to
$\partial/\partial t$ and the closed one-form $L_\mu dx^\mu$ to $dy$,
where $t$ and $y$ are two of the four
coordinates.  In particular, we write down the four-dimensional metric as
\begin{equation}
ds_{4}^2=-h^{-2}(dt+V_idx^i)^2+h^2(dy^2+h_{ij}dx^idx^j),
\label{eq:ds4}
\end{equation}
where $i,j=1,2$. The remaining
components of the metric are $V_i$ and $h_{ij}$, to be determined below,
and $h^2$, given from (\ref{eq:knorm}) to be
\begin{equation}
h^{-2}=e^H(a^2e^G+(b+a\chi)^2e^{-G}).
\label{eq:hexp}
\end{equation}
In addition, for $L=dy$, (\ref{eq:diffh}) yields the constraints
\begin{equation}
\eta=b\partial_ye^H,\qquad\widetilde\eta=-a\partial_ye^H.
\label{eq:ehcon}
\end{equation}
where we still allow for any of these constants $\eta$, $\widetilde\eta$,
$a$ or $b$ to be zero.

Assuming $e^H=y$, which relates $\eta,\tilde\eta$ to $a,b$ according to (\ref{eq:ehcon}), from the supersymmetry  
variation of the gravitino $\delta\lambda_H$, we find
\be
\bigg(\gamma^{\underline 3}(\widetilde\eta^2+(\eta-\chi\widetilde\eta)^2)^{\ft 12}
-(\eta-\chi\widetilde\eta)e^{-G}+i\gamma_5\widetilde\eta\bigg)\epsilon=0.
\label{eq:proj1}
\ee
The projector (\ref{eq:proj1}) is easily solved by 
\be
\epsilon=\exp(i\alpha\gamma_5\gamma^{\underline 3})\epsilon_1,\qquad
\hbox{where~~}
\sinh(2\alpha)=\frac{\tilde\eta e^G}{\eta-\chi\tilde\eta},\qquad \hbox{and~~}
\gamma^{\underline 3}\epsilon_1=\epsilon_1.
\label{eq:killingspinor1}
\ee
The norm of the spinor $\epsilon_1$ is obtained from knowledge of the
spinor bilinears $f_1, f_2$. Choosing a particular representation
of the 4-dimensional Dirac matrices, let's say the chiral representation, 
we compute
\be
\epsilon_1^T = (\epsilon_0, 0, -i\epsilon_0, 0), \qquad \hbox{with~~}
|\epsilon_0|^2=\ft 12 e^{\ft 12(H+G)}\sinh(2\alpha)^{-1}
\ee
and we can set the phase of $\epsilon_0$ to zero (i.e. take $\epsilon_0$ real).

There is another set of spinor bilinears which provides useful information,
namely 
\be
\omega=\epsilon^T C\gamma_\mu\epsilon \;dx^\mu,
\ee
where $C$ is the charge conjugation matrix $\gamma_\mu^T=-C\gamma^\mu C^{-1}$.
One can check that the one-form $\omega$ is closed.
Substituting the Killing spinor (\ref{eq:killingspinor1}) into $\omega$ 
we obtain
\bea
\omega_{\underline 2}=\frac{\cosh(2\alpha)e^{\ft 12(H+G)}}{\sinh(2\alpha)}=\frac1{\widetilde\eta}h^{-1},\qquad
\omega_{\underline 1}=\frac{1}{\widetilde\eta}ih^{-1},
\eea
where we used that for chiral representation $C=i\gamma^2\gamma^0$.
Given that $\omega=h(\omega_{\underline 2} e^{\underline 2}+
\omega_{\underline 1} e^{\underline 1})$ is closed we conclude yet again 
that the 2-dimensional space parameterized by $x^1, x^2$ is 
(conformally) flat.  

We now have sufficient information to fix the form of the
field strengths $F_{(2)}$ as well as $dV$.  For $F_{(2)}$, we use the
component relations
\begin{equation}
F_{\mu\nu}K^\nu=-2a\partial_\mu e^{H+G}-2(b+a\chi)e^{H-G}\partial_\mu\chi,
\qquad
\widetilde F_{\mu\nu}K^\nu=-2(b+a\chi)\partial_\mu e^{H-G}
-2ae^{H-G}\partial_\mu\chi,
\end{equation}
obtained from (\ref{eq:fkeqn}) as well as the explicit form of the
metric (\ref{eq:ds4}) to find
\begin{eqnarray}
F_{(2)}\!&=&\!-2\Bigl[a\,de^{H+G}+(b+a\chi)e^{H-G}d\chi\Bigr]\wedge(dt+V)
-2h^2e^G*_3d\Bigl[(b+a\chi)e^{H-G}\Bigr],\nonumber\\
\widetilde F_{(2)}\!&=&\!-2d\Bigl[(b+a\chi)e^{H-G}\Bigr]\wedge(dt+V)
+2h^2e^{-G}*_3\Bigl[a\,de^{H+G}+(b+a\chi)e^{H-G}d\chi\Bigr],\qquad\quad
\label{eq:f2form}
\end{eqnarray}
where $*_3$ denotes the Hodge dual with respect to the flat spatial metric.
For $dV$, we take the antisymmetric part of $\nabla_\mu K_\nu$ in
(\ref{eq:difdi}), written in form notation as
\begin{equation}
dK=\ft12e^{-\fft{1}2(H+G)}(f_2F_{(2)}-f_1*F_{(2)}),
\end{equation}
and substitute in the expressions for the Killing vector
$K=-h^{-2}(dt+V)$ as well as for $F_{(2)}$.
This gives both the known expression for $h^{-2}$, namely
(\ref{eq:hexp}), as well as the relation
\begin{equation}
dV=-h^4e^H*_3\Bigl[2a(b+a\chi)dG+\Bigl((b+a\chi)^2e^{-2G}-a^2\Bigr)
d\chi\Bigr].
\label{eq:dv}
\end{equation}
Note that when $\chi=0$ this reduces to $dV=-2abh^4e^H*_3dG$, obtained
in \cite{Lin:2004nb,Liu:2004ru}.

Combining the expression for $dV$ in (\ref{eq:dv}) with that of $h^2$
in (\ref{eq:hexp}), we may re-express the one-forms $dG$ and $d\chi$ in
terms of $dV$ and $d(h^2)$.  This allows us to rewrite $F_{(2)}$ in
a more suggestive manner
\begin{eqnarray}
F_{(2)}\!&=&\!-2\Bigl[a\,de^{H+G}+(b+a\chi)e^{H-G}d\chi\Bigr]\wedge(dt+V)
-2ae^{H+G}dV+2(b+a\chi)*_3e^Hd(h^2),\nonumber\\
\widetilde F_{(2)}\!&=\!&-2d\Bigl[(b+a\chi)e^{H-G}\Bigr]\wedge(dt+V)
-2(b+a\chi)e^{H-G}dV-2a*_3e^Hd(h^2).
\label{eq:f2f2t}
\end{eqnarray}
In addition, so long as $a\ne0$, the expression for $dV$ may be written as
\begin{equation}
dV=*_3a^{-1}e^{-H}d\left(\fft{b+a\chi}{a^2e^{2G}+(b+a\chi)^2}\right),
\qquad(a\ne0),
\label{eq:dv1}
\end{equation}
where we have again used the form of $h^2$ in (\ref{eq:hexp}).
For $a=0$, on the other hand, we would instead find simply
\begin{equation}
dV=-b^{-2}e^{-H}*_3d\chi,\qquad(a=0).
\label{eq:dv2}
\end{equation}

The above results have all been obtained as a consequence of the Killing
spinor equations.  However, as is well known, for partial supersymmetry,
the first order Killing spinor equations generally imply only a subset
of the complete equations of motion.  This is indeed the case for the
reduced $\mathcal N=(1,0)$ system.  The relation between the Killing
spinor equations and the bosonic equations of motion is investigated
in Appendix~\ref{sec:apb}.  The result of that analysis indicates that,
so long as the $\widetilde F_{(2)}$ Bianchi identity $d\widetilde F_{(2)}=0$
and equation of motion $dF+\widetilde F\wedge d\chi=0$ are satisfied,
we are then ensured a complete solution to the equations of motion.

Taking an exterior derivative of the expressions in (\ref{eq:f2f2t}),
we obtain
\begin{eqnarray}
d\widetilde F_{(2)}&=&-2a \;d\Bigl[*_3e^Hd(h^2)\Bigr]-2(b+a\chi)e^{H-G}d^2V,
\nonumber\\
dF_{(2)}+\widetilde F_{(2)}\wedge d\chi&=&-2ae^{H+G}d^2V+2(b+a\chi)
\;d\Bigl[*_3e^Hd(h^2)\Bigr].
\end{eqnarray}
Note that $dV$ is not automatically closed; this must be imposed as
an additional consistency condition on either (\ref{eq:dv1}) or
(\ref{eq:dv2}).  We thus see that the bubbling AdS$_3$ analysis leads
to two independent second order equations
\begin{equation}
d\Bigl[*_3e^Hd(h^2)\Bigr]=0,\qquad
d\Bigl[*_3e^{-H}dz\Bigr]=0,
\label{eq:zheom}
\end{equation}
where we have defined
\begin{equation}
z=\fft1{2ab}-\fft{b+a\chi}{a(a^2e^{2G}+(b+a\chi)^2)}
=\fft1{2ab}\left(\fft{a^2(\chi^2+e^{2G})-b^2}{a^2e^{2G}+(b+a\chi)^2}
\right),
\label{eq:zgrel}
\end{equation}
so that
\begin{equation}
dV=-*_3e^{-H}dz.
\end{equation}
Note that this expression remains valid in the limits $a\to0$ or
$b\to0$, provided an (unimportant) infinite constant is subtracted.

The fact that there are now two second order equations, (\ref{eq:zheom}),
indicates that the bubbling AdS$_3\times S^3$ geometries have a
different characteristic from that of the bubbling AdS$_5\times S^5$
solutions of \cite{Lin:2004nb}.

\subsection{Specialization of $\eta$ and $\widetilde\eta$}

So far, we have left $\eta$ and $\widetilde\eta$ unspecified and
performed a general supersymmetry analysis.  We now specialize
the Killing spinor $\mathrm U(1)$ charges, considering the four
possibilities for either of $\eta$ and $\widetilde\eta$ vanishing
or non-vanishing.

\subsubsection{Both $\eta$ and $\widetilde\eta$ non-vanishing}

We begin with the case of both $\eta$ and $\widetilde\eta$ non-vanishing.
To be specific, we take $a=-\widetilde\eta=1$ as well as $b=\eta=1$,
which was chosen to satisfy (\ref{eq:abeta}).  In this case,
(\ref{eq:ehcon}) yields the simple result $e^H=y$, so that (\ref{eq:hexp})
becomes
\begin{equation}
h^{-2}=y(e^G+(1+\chi)e^{-G}).
\label{eq:h11exp}
\end{equation}
We now see that, in the absence
of the axion ($\chi=0$), this expression reduces to that of
\cite{Lin:2004nb,Liu:2004ru}, namely $h^{-2}=2y\cosh G$.  With
$e^H=y$, the second order equations (\ref{eq:zheom}) reduce to
\begin{equation}
d\Bigl[*_3yd(h^2)\Bigr]=0,\qquad
d\Bigl[*_3y^{-1}dz\Bigr]=0.
\label{eq:harmeom}
\end{equation}
The first equation is a new one compared with the AdS$_5\times S^5$ case,
and indicates that $h^2$ is harmonic in a four-dimensional auxiliary space
$\mathbb R^2\times\mathbb R^2$, restricted to $s$-waves only in the second
$\mathbb R^2$.  The second equation, on the other hand, is a direct
generalization of the expression for $z$ introduced in \cite{Lin:2004nb}.
Thus $z/y^2$ is harmonic in a six-dimensional auxiliary space
$\mathbb R^2\times\mathbb R^4$, restricted to $s$-waves in the $\mathbb R^4$.
In contrast with \cite{Lin:2004nb}, however, the relation between $z$
and $G$ is now given by (\ref{eq:zgrel}), and reads
\begin{equation}
z=\fft12\fft{e^{2G}-(1-\chi^2)}{e^{2G}+(1+\chi)^2},
\label{eq:z11exp}
\end{equation}
which generalizes the expression $z=\fft12\tanh G$ for a non-vanishing
axion.

Note that the introduction of the axion has removed the
$dG\wedge*_3dG=dH\wedge*_3dH$ that was identified in \cite{Liu:2004ru}.
This, however, comes at the expense of introducing a second harmonic
function to the bubbling AdS$_3$ construction.

To summarize, the bubbling AdS$_3\times S^3$ solution is given as:
\begin{eqnarray}
ds_6^2 &=& -h^{-2}(dt+V_idx^i)^2
+h^2(dy^2+\delta_{ij}dx^idx^j)+y\Bigl[e^Gd\phi_1^2
+e^{-G}(d\phi_2+\chi d\phi_1)^2\Bigr],\qquad\nonumber\\
\widetilde F_{(2)} &=&-2\left[
d\left((1+\chi)ye^{-G}\right)\wedge (dt+V)
-h^2e^{-G}*_3\left(d(ye^G)+(1+\chi)ye^{-G}d\chi\right)\right],
\label{eq:bubbling}
\end{eqnarray}
where
\begin{equation}
h^{-2} = y(e^G+(1+\chi)e^{-G}),\qquad z=\fft12
\fft{e^{2G}-(1-\chi^2)}{e^{2G}+(1+\chi)^2},\qquad
dV = -\ft{1}{y}*_3dz.
\end{equation}
The functions $h^2$ and $z$ must satisfy the harmonic equations
(\ref{eq:harmeom}).

\subsubsection{Only $\eta$ non-vanishing}

With the introduction of the axion, the spinor $\mathrm U(1)$ charges
$\eta$ and $\widetilde\eta$ are no longer interchangeable.  Here we consider
$\eta=1$ and $\widetilde\eta=0$.  In this case, the constraint
(\ref{eq:abeta}) indicates that $a=0$.  Avoiding the degenerate situation,
we now take $b=\eta=1$, so that (\ref{eq:ehcon}) again gives $e^H=y$.  This
time, however, the relation (\ref{eq:hexp}) yields a single exponential,
$h^{-2}=ye^{-G}$, while (\ref{eq:zgrel}) gives simply $z=\chi$ (after
removing an unimportant infinite constant).  In addition,
the field strength $F_{(2)}$ is given by (\ref{eq:f2form})
\begin{equation}
\widetilde F_{(2)}=-2\,d(ye^{-G})\wedge(dt+V)+2e^{-G}*_3d\chi.
\end{equation}

To ensure a solution of the equations of motion, we must also satisfy
the second order equations
\begin{equation}
d\Bigl[*_3yd(h^2)\Bigr]=0,\qquad
d\Bigl[*_3y^{-1}d\chi\Bigr]=0.
\end{equation}
As a result, the solution may be written as
\begin{eqnarray}
ds_6^2&=&{\cal H}^{-1}(-(dt+V_idx^i)^2+d(\phi_2+\chi d\phi_1)^2)+
{\cal H}(\delta_{ij}dx^idx^j+dy^2+y^2d\phi_1^2),\nonumber\\
\widetilde F_{(2)}&=&2(dt+V)\wedge d({\cal H}^{-1})-2{\cal H}^{-1}dV,\qquad
dV=-\fft1y*_3d\chi,
\label{eq:multicenter}
\end{eqnarray}
where we have introduced the four-dimensional harmonic function
${\cal H}=h^2=\fft1ye^G$.  This generalizes the familiar multi-centered
string solution in six-dimensions (which is obtained by taking $\chi=0$),
restricted to singlet configurations along the $\phi_1$ direction,
assuming that the $S^1$ parameterized by $\phi_2$ has decompactified.
Turning on the axion (which also turns on $V$) yields more general 1/2~BPS
solutions of the form obtained in \cite{GMR}.

\subsubsection{Only $\widetilde\eta$ non-vanishing}

With $\eta=0$ and $\widetilde\eta=-1$, the constraint
(\ref{eq:abeta}) indicates that $b=0$.  Setting $a=-\widetilde\eta=1$,
we once again see that $e^H=y$.  Hence the solutions obtained in this
fashion also satisfy (\ref{eq:harmeom}), and thus fall in the same
class.  In particular, (\ref{eq:hexp}) and (\ref{eq:zgrel}) gives
\begin{equation}
h^{-2}=ye^G(1+e^{-2G}\chi^2),\qquad
z=-\fft{e^{-2G}\chi}{1+e^{-2G}\chi^2}
\end{equation}
(where again an unimportant constant was removed from $z$).

In fact, these expressions are readily obtained from the previous case
of $\eta=1$, $\widetilde\eta=0$ by performing the $\mathrm{SL}(2,\mathbb Z)$
transformation $\tau\to-1/\tau$ with the identification
\begin{equation}
h^2=\fft1y\Im\tau,\qquad z=\Re\tau.
\end{equation}
In particular, for $\tau=\chi+ie^G$, we see that
\begin{equation}
\tau\to-\fft1\tau=\fft{-e^{-2G}\chi+ie^{-G}}{1+e^{-2G}\chi^2}.
\end{equation}
Note, also, that the transformation
\begin{equation}
\tau\to\tau+1\to-\fft1{\tau+1}
=\fft{-e^{-2G}(1+\chi)+ie^{-G}}{1+e^{-2G}(1+\chi)^2}
\end{equation}
relates the $\eta=1$, $\widetilde\eta=0$ solution to the (two charge)
$\eta=1$, $\widetilde\eta=-1$ case.  In other words, the two $\mathrm U(1)$
charges naturally form a two-component $\mathrm{SL}(2,\mathbb Z)$ charge
vector $(\eta,\widetilde\eta)$, and all three examples
$(b,a)=(\eta,-\widetilde\eta)=(1,1)$, $(1,0)$ and $(0,1)$ fall into the
same $\mathrm{SL}(2,\mathbb Z)$ conjugacy class.

\subsubsection{Both $\eta$ and $\widetilde\eta$ vanishing}

Finally, the case $\eta=\widetilde\eta=0$ is distinct from the previous
ones, as it corresponds to a standard Kaluza-Klein reduction with
uncharged Killing spinors.  In this case, the constraint (\ref{eq:abeta})
becomes trivial, so that $a$ and $b$ may take on arbitrary values.
While $(\eta,-\widetilde\eta)=(0,0)$ is a $\mathrm{SL}(2,\mathbb Z)$ singlet,
we assume that at least one of $a$ or $b$ is non-vanishing, so that
$(b,a)$ remains a $\mathrm{SL}(2,\mathbb Z)$ doublet.  In this case,
(\ref{eq:ehcon}) implies that $H$ is a constant, which we take to be zero.

Up to a $\mathrm{SL}(2,\mathbb Z)$ transformation, we take the simplest
case $(b,a)=(1,0)$.  For this case, and with $H=0$, (\ref{eq:hexp}) 
and (\ref{eq:zgrel}) gives
\begin{equation}
h^{-2}=e^{-G},\qquad z=\chi,
\end{equation}
and (\ref{eq:f2form}) yields
\begin{equation}
\widetilde F_{(2)}=-2d(e^{-G})\wedge(dt+V)-2e^{-G}dV,
\end{equation}
with $dV=-*_3d\chi$.  In this case, the solution has the form
\begin{eqnarray}
ds_6^2&=&{\cal H}^{-1}(-(dt+V_idx^i)^2+d(\phi_2+\chi d\phi_1)^2)+
{\cal H}(\delta_{ij}dx^idx^j+dy^2+d\phi_1^2),\nonumber\\
\widetilde F_{(2)}&=&2(dt+V)\wedge d({\cal H}^{-1})-2{\cal H}^{-1}dV,\qquad
dV=-*_3d\chi,
\end{eqnarray}
where $\mathcal H=h^2=e^G$.  Note that here the equations of motion
are
\begin{equation}
d*_3d\mathcal H=0,\qquad d*_3d\chi=0,
\end{equation}
so that both ${\cal H}$ and $\chi$ are harmonic in $\mathbb R^3$
spanned by $(x^1,x^2,y)$.  This solution is in fact of the same
form as (\ref{eq:multicenter}), and, in the limit of vanishing axion,
represents a multi-centered string solution smeared out along the
$\phi_1$ direction.  Note that here both circles have decompactified.

%%%%%%%%%%%%%%%%%%%%%%%%%%%%%%%%%%%%%%%%%%%%%%%%%%%%%%%%%%%%%%%%%%%%%%%%%%%%%
\section{Discussion}

We have constructed a family of 1/2 BPS solutions of minimal six-dimensional supergravity.
These solutions inherit the SL$(2,\mathbb R)$/U(1) isometries of the $T^2$ reduction ansatz.
The complex structure is parameterized by $\tau=\chi+i e^G$, whereas the volume of 
$T^2$ is given by $e^H$. We have thus generalized our previous $S^1\times S^1$ reduction ansatz,
with the radii of the two circles given by $e^{H+G}$ and $e^{H-G}$, by allowing for
a non-vanishing axion. The $S^1\times S^1$ solutions were written in terms of a harmonic
function on an auxiliary six dimensional space $\mathbb R^2\times \mathbb R^4$, just
as it was the case for the $S^3\times S^3$ reduction of type IIB supergravity.
However, the $S^1\times S^1$ reduction turned out to be inconsistent, due to an
additional constraint that the four dimensional gauge field had to satisfy:
$F_{(2)}\wedge F_{(2)}=0$.
Moreover, this additional constraint translated into another non-linear differential equation
which the harmonic function had to obey. This ultimately prohibited the existence of a family of solutions, 
even though few isolated solutions were found, such as AdS$_3\times S^3$, the maximally symmetric 
plane wave, and the multi-center string, provided that the six dimensional Killing spinors
were carrying some momentum on the two $S^1$.  

The effect of adding the axion among the Kaluza-Klein states to be kept in the reduction
is to remove the constraint rendering the bosonic reduction consistent. 
At the same time, the 1/2 BPS six dimensional solutions are characterized by {\it two} functions, 
one being, as before, harmonic on the auxiliary six-dimensional space $\mathbb R^2\times \mathbb R^4$, 
while the other being harmonic on a four dimensional auxiliary space $\mathbb R^2\times\mathbb R^2$.
This brings a distinct flavor to the 1/2 BPS solutions of minimal six dimensional
supergravity (with two U(1) isometries) in comparison to the 1/2 BPS
solutions of type IIB supergravity (with SO(4)$\times$SO(4) isometry).

We have also explicitly constructed the Killing spinors associated with 
the six-dimensional solutions. The Killing spinors are again charged under the two U(1)
isometries, but this time their U(1) charges combine into an SL$(2,\mathbb Z)$ doublet, and 
their corresponding solutions are mapped into each other under the action
of SL$(2,\mathbb Z)$. 

In fact, the solutions we have found appear to be a particular case
of a larger class of six dimensional D1-D5 solutions with angular momentum 
obtained by Lunin, Maldacena and Maoz \cite{lmm} (which in turn are
desingularized versions of those constructed in \cite{Lunin:2002bj}).
This is most transparent if we choose to compare one of our solutions
(\ref{eq:multicenter}), corresponding to the U(1) charges
$\widetilde\eta =0$ and $\eta\neq 0$, to the solution (2.1) in \cite{lmm}. 
A brief inspection of (2.1) in \cite{lmm} reveals that for solutions of minimal
six dimensional supergravity we should identify the functions $f_1$ and $f_5$, meaning we 
must enforce $\dot {\vec F}(v) \dot{\vec F}(v)=1$.
The dictionary between our (\ref{eq:multicenter}) and  (2.1) 
in \cite{lmm} includes  ${\cal H}\rightarrow f_1, V\rightarrow A_i dx^i, \chi d\phi_1 \rightarrow B_i dx^i,
 \{x^1,x^2,y,\phi_1\}\rightarrow \{\vec x\}$ and $\phi_2\rightarrow y$.  
Our solutions have also an additional Killing vector, namely $\partial_{\phi_1}$.
This restricts the profile $\vec F(\vec x(v))$ dependence to $\vec F(x^1 (v), x^2(v))$ at $y=0$.
The solution (2.1) in \cite{lmm}, which was derived by applying a chain of dualities 
to a fundamental string carrying momentum, was shown to be regular provided that the profile
of the fundamental string, specified by $\vec F(v)$, obeyed a few conditions:
it was not self-intersecting, and $|\dot{\vec F}(v)|\neq 0$.

The main outcome of this comparison between our solution (\ref{eq:multicenter})
and (2.1) of \cite{lmm} is that we realize that subsequent regularity conditions
will relate the boundary conditions of our two harmonic functions: $\chi/y^2$ and 
$h^2=\cal H$, since equation (2.2) of \cite{lmm} can be rewritten in terms of Green's
function associated with our harmonic functions. 
Therefore the bubbling picture for AdS$_3\times S^3$ is completed upon enforcing 
regularity, and the two dimensional droplets of the bubbling AdS$_5\times S^5$
solution have morphed into boundaries specified by the profile $\vec F(v)$.    
Understanding the regularity properties of our solutions and their 
direct relationship with the chiral primaries of the dual CFT deserves further study.
It would also be desirable to understand the peculiarities of the 
giant gravitons (their unrestricted growth, their discrete angular momenta) 
in terms of the bubbling $AdS_3$ picture.

%%%%%%%%%%%%%%%%%%%%%%%%%%%%%%%%%%%%%%%%%%%%%%%%%%%%%%%%%%%%%%%%%%%%%%%%%%%%%%
\section*{Note added}
While this work was under completion, we became aware of
\cite{Martelli:2004xq}, where bubbling AdS$_3\times S^3$ solutions of
the form (\ref{eq:bubbling}) were also obtained.  The analysis of
\cite{Martelli:2004xq} followed directly from the complete six-dimensional
classification of \cite{GMR} by choosing an appropriate reduction with
three commuting Killing symmetries $\partial/\partial x^+$,
$\partial/\partial x^-$ and $\partial/\partial\phi$.  From a
six-dimensional point of view, the Killing vector $K^M=
\bar\epsilon\Gamma^M\epsilon$ is null, leading to a natural $x^+$, $x^-$
basis.  To make the comparison more direct, we may invert the expressions
(\ref{eq:h11exp}) and (\ref{eq:z11exp}) to obtain
\begin{equation}
e^{-G}=h^2y+(h^2y)^{-1}(z-\ft12)^2,\qquad
\chi=-\fft{h^2y+(h^2y)^{-1}(z^2-\fft14)}{h^2y+(h^2y)^{-1}(z-\fft12)^2},
\end{equation}
which correspond to the metric elements given in \cite{Martelli:2004xq}.

%% Our work, on the other hand, extends the techniques 
%% of \cite{Lin:2004nb},\cite{Liu:2004ru},\cite{Gauntlett:2002sc},
%% \cite{Gauntlett:2002nw} to the $T^2$ torus reduction of minimal 6 
%% dimensional supergravity.

Furthermore, the work of \cite{Martelli:2004xq} demonstrates that
the bubbling AdS$_3\times S^3$ solutions are in fact a restricted sub-class
of {\it all} the 1/2~BPS solutions of \cite{GMR}.  This brings up an
appropriate note of caution, namely that the bubbling forms of 1/2~BPS
solutions are not necessarily exhaustive, as far as the full theory is
concerned, but only correspond to sub-classes where additional
Killing symmetries are imposed on the background.

%%%%%%%%%%%%%%%%%%%%%%%%%%%%%%%%%%%%%%%%%%%%%%%%%%%%%%%%%%%%%%%%%%%%%%%%%%%%%%
\section*{Acknowledgments}
We wish to thank W.Y.~Wen for early discussions on obtaining bubbling
AdS$_3\times S^3$ solutions via $T^2$ reductions.  This work was supported
in part by the US~Department of Energy under grant DE-FG02-95ER40899.
We are thankful to Oleg Lunin for bringing the paper \cite{lmm} to our attention.

%%%%%%%%%%%%%%%%%%%%%%%%%%%%%%%%%%%%%%%%%%%%%%%%%%%%%%%%%%%%%%%%%%%%%%%%%%%%%%
\appendix

%%%%%%%%%%%%%%%%%%%%%%%%%%%%%%%%%%%%%%%%%%%%%%%%%%%%%%%%%%%%%%%%%%%%%%%%%%%%%
\section{Integrability of the Killing spinor equations}
\label{sec:apb}

In \cite{Liu:2004ru}, the integrability of the supersymmetry variations
(\ref{eq:6to4susy}) was obtained in the absence of the axion.  The
results of that work is easily extended to the present case.  We take
\begin{equation}
\delta\psi_\mu={\cal D}_\mu\epsilon,\qquad
\delta\lambda_H=\Delta_H,\qquad\delta\lambda_G=\Delta_G,
\end{equation}
where, from (\ref{eq:6to4susy}), we read off
\begin{eqnarray}
{\cal D}_\mu&=&\nabla_\mu-\ft{i}4\gamma^5e^{-G}\partial_\mu\chi
+\ft{i}{16}e^{-\fft12(H+G)}F_{\nu\lambda}
\gamma^{\nu\lambda}\gamma_\mu,\nonumber\\
\Delta_H&=&\gamma^\mu\partial_\mu H+e^{-\fft12H}((\eta-\chi\widetilde\eta)
e^{-\fft12G}-i\widetilde\eta\gamma_5e^{\fft12G}),\nonumber\\
\Delta_G&=&\gamma^\mu\partial_\mu G+i\gamma^5\gamma^\mu e^{-G}\partial_\mu\chi
-\ft{i}4e^{-\fft12(H+G)}F_{\mu\nu}\gamma^{\mu\nu}
+e^{-\fft12H}((\eta-\chi\widetilde\eta)e^{-\fft12G}
+i\widetilde\eta\gamma_5e^{\fft12G}).\nonumber\\
\label{eq:ddsusy}
\end{eqnarray}
Here we may read off three independent integrability conditions, related
to the commutators $[\mathcal D_\mu,\mathcal D_\nu]$, $[\mathcal D_\mu,
\Delta_H]$ and $[\mathcal D_\mu,\Delta_G]$.
For $[{\cal D}_\mu,{\cal D}_\nu]$, we obtain
\begin{eqnarray}
\gamma^\mu[{\cal D}_\mu,{\cal D}_\nu]&=&
\ft12[R_{\nu\sigma}-\ft14e^{-(H-G)}(\widetilde F^2{}_{\nu\sigma}-
\ft14g_{\mu\nu}\widetilde F^2)\nonumber\\
&&\qquad
-\ft12(\partial_\nu H\partial_\sigma H+\partial_\nu G\partial_\sigma G
+e^{-2G}\partial_\nu\chi\partial_\sigma\chi)
-\nabla_\nu\nabla_\sigma H]\gamma^\sigma\nonumber\\
&&+\ft{i}{16}e^{-\fft12(H+G)}(\partial_{[\mu}F_{\lambda\sigma]}
+\widetilde F_{[\mu\lambda}\partial_{\sigma]}\chi)
\gamma^{\mu\lambda\sigma}\gamma_\nu+\ft{i}8e^{-\fft12(H-G)}
\nabla^\mu(e^{-G}F_{\mu\lambda})\gamma^\lambda\gamma_\nu\nonumber\\
&&+\ft12[{\cal D}_\nu,\Delta_H]+\ft18\partial_\nu(H+G)(\Delta_H+\Delta_G)
\nonumber\\
&&+\ft18[\partial_\nu(H-G)-\ft{i}{4}e^{-\fft12(H+G)}F_{\lambda\sigma}
\gamma^{\lambda\sigma}\gamma_\nu+2i\gamma^5e^{-G}\partial_\nu\chi]
(\Delta_H-\Delta_G).
\label{eq:caldd}
\end{eqnarray}
Note that we have used the relation $F_{(2)}=e^G*_4\widetilde F_{(2)}$
to rewrite the Einstein equation in terms of $\widetilde F_{(2)}$.
Since the last two lines above vanish on Killing spinors, this integrability
condition yields the Einstein equation in conjunction with the Bianchi
identity and equation of motion for $\widetilde F_{(2)}$.

Turning to the $[\mathcal D_\mu,\Delta_H]$ condition, we find
\begin{eqnarray}
\gamma^\mu[{\cal D}_\mu,\Delta_H]&=&\square H+\partial H^2\nonumber\\
&&-[\gamma^\mu\partial_\mu H-\ft{i}2\gamma^5\gamma^\mu e^{-G}\partial_\mu
\chi+\ft{i}8e^{-\fft12(H+G)}F_{\mu\nu}\gamma^{\mu\nu}\nonumber\\
&&\kern6em-\ft12e^{-\fft12H}((\eta-\chi\widetilde\eta)e^{-\fft12G}
+i\widetilde\eta\gamma_5e^{\fft12G})]\Delta_H\nonumber\\
&&-\ft12e^{-\fft12H}((\eta-\chi\widetilde\eta)e^{-\fft12G}
-i\widetilde\eta\gamma_5e^{\fft12G})\Delta_G.
\label{eq:caldh}
\end{eqnarray}
On Killing spinors this yields precisely the $H$ equation of motion,
$\nabla^\mu\left(e^H\nabla_\mu H\right)=0$, of (\ref{eq:aeeom}).  This
indicates that the $H$ equation of motion (and hence the solution for $H$)
is guaranteed by supersymmetry.

Finally, the $[\mathcal D_\mu,\Delta_G]$ integrability condition becomes
\begin{eqnarray}
\gamma^\mu[{\cal D}_\mu,\Delta_G]&=&\square G+\partial H\partial G
+e^{-2G}(\partial\chi)^2-\ft18e^{-(H-G)}\widetilde F^2\nonumber\\
&&-i\gamma^5e^{-G}[\square\chi+\partial(H-2G)\partial\chi
+\ft1{16}e^{-H+2G}\epsilon_{\mu\nu\lambda\sigma}\widetilde F_{\mu\nu}
\widetilde F_{\lambda\sigma}]\nonumber\\
&&-\ft{i}4e^{-\fft12(H+G)}(\partial_{[\mu}F_{\nu\lambda]}
+\widetilde F_{[\mu\nu}\partial_{\lambda]}\chi)
\gamma^{\mu\nu\lambda}-\ft{i}2e^{-\fft12(H-G)}\nabla^\mu(e^{-G}F_{\mu\nu})
\gamma^\nu\nonumber\\
&&-\ft12[\gamma^\mu\partial_\mu G-i\gamma^5\gamma^\mu e^{-G}\partial_\mu\chi]
\Delta_H
-\ft12[\gamma^\mu\partial_\mu H+i\gamma^5\gamma^\mu e^{-G}\partial_\mu\chi]
\Delta_G.
\label{eq:caldg}
\end{eqnarray}
In addition to the Bianchi identity and equation of motion for
$\widetilde F_{(2)}$, this condition yields the equations of motion for
the $\mathrm{SL}(2,\mathbb R)$ scalar $\tau=\chi+ie^G$.  In the absence
of an axion, it is this equation that leads to the
$\widetilde F_{(2)}\wedge \widetilde F_{(2)}=0$ constraint of
\cite{Liu:2004ru}.

Disregarding the $H$ equation, which is automatically satisfied on a
supersymmetric background, we see that the existence of a Killing spinor
only ensures that linear combinations of the Einstein equation,
$\widetilde F_{(2)}$ Bianchi identity and equation of motion, and $\tau$
equation of motion are satisfied.  Although somewhat more care is
needed to fully disentangle the bosonic equations of motion in
(\ref{eq:caldd}) and (\ref{eq:caldg}), we see that, so long as the
$\widetilde F_{(2)}$ Bianchi identity and equation of motion are
satisfied, (\ref{eq:caldd}) then guarantees that the Einstein equation
will hold, and further (\ref{eq:caldg}) will ensure the full $\tau$ equation
of motion (so long as the Killing spinor has indefinite $\gamma^5$
chirality).  We thus conclude that, for obtaining supersymmetric backgrounds,
it would be sufficient to satisfy the $\widetilde F_{(2)}$ Bianchi identity
and equation of motion in addition to the Killing spinor equations themselves.

%%%%%%%%%%%%%%%%%%%%%%%%%%%%%%%%%%%%%%%%%%%%%%%%%%%%%%%%%%%%%%%%%%%%%%%%%%%%%%
\section{Differential identities for the spinor bilinears}
\label{sec:apc}

The supersymmetric construction of
\cite{Gauntlett:2002sc,Gauntlett:2002nw,GMR,Gauntlett:2004zh} proceeds
by postulating the existence of a Killing spinor $\epsilon$ and then
forming the tensors $f_1$, $f_2$, $K_\mu$, $L_\mu$ and $Y_{\mu\nu}$ from
spinor bilinears (\ref{eq:bilinear}).  The algebraic identities of interest
were given in the text in (\ref{eq:h2}).  Here we tabulate the differential
identities obtained by demanding that $\epsilon$ solves the Killing
spinor equations obtained from (\ref{eq:6to4susy}).

First, by assuming $\delta\psi_\mu=0$, we may demonstrate that
\begin{eqnarray}
\nabla_\mu f_1&=&\ft14e^{-\fft{1}{2}(H+G)}*F_{\mu\nu}K^\nu+\ft 12 f_2 e^{-G}\partial_\mu\chi,
\nonumber\\
\nabla_\mu f_2&=&-\ft14e^{-\fft{1}{2}(H+G)}F_{\mu\nu}K^\nu-\ft 12 f_1e^{-G}\partial_\mu\chi
,\nonumber\\
\nabla_\mu K_\nu&=&\ft14e^{-\fft{1}{2}(H+G)}(f_2F_{\mu\nu}-f_1*F_{\mu\nu}),
\nonumber\\
\nabla_\mu L_\nu&=&\ft14e^{-\fft{1}{2}(H+G)}
(\ft12g_{\mu\nu}F_{\lambda\rho}Y^{\lambda\rho}-2F_{(\mu}{}^\lambda
Y_{\nu)\lambda}),\nonumber\\
\nabla_\mu Y_{\nu\lambda}&=&\ft14e^{-\fft{1}2(H+G)}(2g_{\mu[\nu}
F_{\lambda]\rho}L^\rho-2F_{\mu[\nu}L_{\lambda]}+F_{\nu\lambda}L_\mu).
\label{eq:difdi}
\end{eqnarray}
In particular, the equation for $K_\mu$ indicates that $K_{(\mu;\nu)}=0$,
so that $K^\mu$ is Killing.  This is in fact a generic feature of
constructing a Killing vector from Killing spinors.

In addition, the $\delta\chi_H=0$ condition allows us to derive the
additional relations
\begin{eqnarray}
&&K^\mu\partial_\mu H=0,\kern9.3em
(\eta-\chi\widetilde\eta) f_2=-\widetilde\eta e^Gf_1,\nonumber\\
&&L^\mu\partial_\mu H=(\eta-\chi\widetilde\eta) e^{-\fft12(H+G)}f_1-\widetilde\eta
e^{-\fft12(H-G)}f_2,\nonumber\\
&&(\eta-\chi\widetilde\eta) e^{-\fft12(H+G)}L_\mu=f_1\partial_\mu H,\kern4.3em
\widetilde\eta e^{-\fft12(H-G)}L_\mu=-f_2\partial_\mu H,\nonumber\\
&&(\eta-\chi\widetilde\eta) e^{-\fft12(H+G)}K_\mu=*Y_\mu{}^\nu\partial_\nu H,\kern3em
\widetilde\eta e^{-\fft12(H-G)}K_\mu=Y_\mu{}^\nu\partial_\nu H,\nonumber\\
&&2L_{[\mu}\partial_{\nu]}H=0,\kern3em
2K_{[\mu}\partial_{\nu]}H=(\eta-\chi\widetilde\eta) e^{-\fft12(H+G)}*Y_{\mu\nu}
+\widetilde\eta e^{-\fft12(H-G)}Y_{\mu\nu}.
\label{eq:difhi}
\end{eqnarray}
Similarly, the $\delta\chi_G=0$ condition yields the relations
\begin{eqnarray}
&&K^\mu\partial_\mu G=0,\kern10em
\ft14 F_{\mu\nu}*Y^{\mu\nu}=(\eta-\chi\tilde\eta) f_2-\widetilde\eta e^Gf_1,
\nonumber\\
&&L^\mu\partial_\mu G=(\eta-\chi\widetilde\eta) e^{-\fft12(H+G)}f_1+\widetilde\eta e^{-\fft12(H-G)}
f_2-\ft14e^{-\fft{1}2(H+G)}F_{\mu\nu}Y^{\mu\nu},\nonumber\\
&&(\eta-\chi\widetilde\eta) e^{-\fft12(H+G)}L_\mu=f_1\partial_\mu G+\ft12e^{-\fft{1}2(H+G)}
*F_{\mu\nu}K^\nu,\nonumber\\
&&\kern4em
\widetilde\eta e^{-\fft12(H-G)}L_\mu=f_2\partial_\mu G+\ft12e^{-\fft{n}2(H+G)}
F_{\mu\nu}K^\nu,\nonumber\\
&&(\eta-\chi\widetilde\eta e^{-\fft12(H+G)}K_\mu=*Y_\mu{}^\nu\partial_\nu G+\ft12e^{-\fft{1}2(H+G)}
*F_{\mu\nu}L^\nu,\nonumber\\
&&\kern4em
\widetilde\eta e^{-\fft12(H-G)}K_\mu=-Y_\mu{}^\nu\partial_\nu G
+\ft12e^{-\fft{1}2(H+G)}F_{\mu\nu}L^\nu,\nonumber\\
&&2L_{[\mu}\partial_{\nu]}G=2e^{-\fft{1}2(H+G)}F_{[\mu}{}^\rho Y_{\nu]\rho},
\nonumber\\
&&2K_{[\mu}\partial_{\nu]}G=(\eta-\chi\widetilde\eta) e^{-\fft12(H+G)}*Y_{\mu\nu}
-\widetilde\eta e^{-\fft12(H-G)}Y_{\mu\nu}-\ft12e^{-\fft{1}2(H+G)}
(f_1*F_{\mu\nu}+f_2F_{\mu\nu}).\nonumber\\
\label{eq:difgi}
\end{eqnarray}
Although the above identities are algebraic and not differential
on the spinor bilinears, they originate from the supersymmetry variations
along the internal directions of the Kaluza-Klein reduction.  So in this
sense, they form a generalized set of `differential identities'.  However,
as they are only algebraic, they prove extremely useful in determining
much of the geometry, as is evident from the analysis of \cite{Lin:2004nb}.

%%%%%%%%%%%%%%%%%%%%%%%%%%%%%%%%%%%%%%%%

\end{document}